\documentclass[aps,showpacs,preprintnumbers,amsmath,amssymb,nofootinbib]{revtex4}

\usepackage{graphicx}
\usepackage{color}
      
\def \be  {\begin{equation}}
\def \ee  {\end{equation}}
\def \ee  {\end{equation}}
\def \bea {\begin{eqnarray}}
\def \eea {\end{eqnarray}}

\begin{document}

\preprint{ECTP-2012-05}

\title{Lorentz Invariance Violation and Generalized Uncertainty Principle}

\author{Abdel Nasser Tawfik\footnote{http://atawfik.net/}}
\affiliation{Egyptian Center for Theoretical Physics (ECTP), Modern University for Technology and Information (MTI), 11571Cairo, Egypt}
\affiliation{World Laboratory for Cosmology And Particle Physics (WLCAPP), 11571 Cairo, Egypt}
\author{H. Magdy}
\affiliation{Egyptian Center for Theoretical Physics (ECTP), Modern University for Technology and Information (MTI), 11571 Cairo, Egypt}
\author{A. Farag Ali}
\affiliation{Physics Department, Faculty of Science, Benha University, Benha 13518, Egypt}

\date{\today}

\begin{abstract}

There are several theoretical indications that the quantum gravity approaches may have predictions for a minimal measurable length, and a maximal observable momentum and throughout a generalization for Heisenberg uncertainty principle. The generalized uncertainty principle (GUP) is based on a momentum-dependent modification in the standard dispersion relation which is conjectured to violate the principle of Lorentz invariance.  From the resulting Hamiltonian, the velocity and time of flight of relativistic distant particles at Planck energy can be derived.  A first comparison is made with recent observations for Hubble parameter in redshift-dependence in early-type galaxies. We find that LIV has two types of contributions to the time of flight delay $\Delta t$ comparable with that observations. Although the wrong OPERA measurement on faster-than-light muon neutrino anomaly, $\Delta t$, and the relative change in the speed of muon neutrino $\Delta v$ in dependence on redshift $z$ turn to be wrong, we utilize its main features to estimate $\Delta v$. Accordingly, the results could not be interpreted as LIV. A third comparison is made with the ultra high-energy cosmic rays (UHECR). It is found that an essential ingredient of the approach combining string theory, loop quantum gravity, black hole physics and doubly spacial relativity and the one assuming a perturbative departure from exact Lorentz invariance.  Fixing the sensitivity factor and its energy dependence are essential inputs for a reliable confronting of our calculations to UHECR. The sensitivity factor is related to the special time of flight delay and the time structure of the signal. Furthermore, the upper and lower bounds to the parameter, $\alpha$ that characterizes the generalized uncertainly principle, have to be fixed in related physical systems such as the gamma rays bursts.

\end{abstract}


\maketitle


\section{Introduction}
\label{sec:intr}

The combination of Heisenberg’s uncertainty principle and finiteness of speed of light $c$ is assumed to lead to creation and annihilation processes, especially when studying Compton wavelength of the particle of interest \cite{garay}. Another consequence of the space-time foamy structure at small scales is the Lorentz invariance violation (LIV). For completeness, we mention that the foamy structure at short distances combines quantum mechanics with general relativity. Different approaches for quantum gravity \cite{guntg}, the yet-to-be-built quantum theory of gravity, have been proposed \cite{garay,isham1}. They provide a set of predictions for a minimal measurable length, and a maximal observable momentum and throughout an essential modification in Heisenberg uncertainty principle. The corresponding effective quantum mechanics would be based on generalized uncertainty principle (GUP) \cite{gupqm}.  According to string theory, loop quantum gravity and black hole physics, GUP is found proportional to a quadratic momenta \cite{refff2b}. Based on doubly spacial relativity, a proportionality to first order moments (linear) has been suggested. As introduced in Ref. \cite{afa1,afa2}, both approaches can be integrated. The resulting one is obviously consistent with doubly special relativity (linear momenta) string theory and black hole physics (quadratic momenta). In this regards, people think to combine quantum mechanics and special relativity and hope to reveal serious difficulties in describing the one-particle theories.  The quantum field theory (QFT) is a perfectly well-defined theoretical framework involving renormalization. Wilson and Weinberg and others taught us that there is nothing wrong with it, i.e. QFT provides precise predictions that are successfully tested in experiments.

The roots of LIV are originated in the suggestion that Lorentz invariance (LI) may represent an approximate symmetry of nature which dates back to about four decades \cite{LI1}. A self-consistent framework for analyzing possible violation of LI was suggested by Coleman and Glashow \cite{cg1,stecker1}. In gamma ray bursts (GRB), the energy-dependent time offsets are investigated in different energy bands  assuming standard cosmological model \cite{jellis1}. A kind of weak indication for the redshift dependence of the time delays suggestive of LIV has been found. A comprehensive review on the main theoretical motivations and observational constraints on Planck scale suppressed Lorentz invariance violation is given in Ref, \cite{revw} and the references therein. Recently, the Planck scale itself turns to be accessible in quantum optics \cite{nature2012}. 
  
Various implications of GUP have been studied so far. Refs. \cite{Sabine,myREV} give very recent reviews. Effects of GUP on atomic, condensed matter systems,  quark gluon plasma, preheating phase and inflationary era of the Universe, black holes production at LHC 
\cite{dvprl,Ali:2010yn,dvcjp,afa2,Elmashad:2012mq,Chemissany:2011nq,Tawfik:2012he,Ali:2012mt} have been investigated. The implications on Saleker-Wigner inequality, compact stars and modified Newton's law of gravitation have been reported \cite{Tawfik:2013uza,Ali:2013ii,Ali:2013ma}. 

The present paper discusses GUP that potentially leads to observable experimental effects related to the violation of Lorentz invariance. Computations in a model characterized by linear modifications is presented and the results are compared with some experimental results. Following the proposal of utilizing astrophysical objects to search for the energy-dependent time of the arrival delays \cite{nature1}, we present an estimation for the time of flight delays and the relative change in the velocity. We compare the results with the observations of Hubble parameter in early-type galaxies in redshift-dependence in section \ref{earlyGal}. Also, we compare the results of muon neutrino based on GUP-approach in section \ref{operra}.  Section \ref{sec:uhecr} is devoted to the calculations which are confronted with the ultra high-energy cosmic rays (UHECR) observations. The conclusions are addressed in section \ref{sec:conl}.


\section{The Approach}
\label{sec:apprch}
 
According to GUP-approach, the momentum of a particle with mass $M$ having distant origin and an energy scale comparable to the Planck's one would be a subject of a tiny modification \cite{afa1,afa2} so that the comoving momenta can be given as
\bea
p_{\nu} = p_{\nu} \left(1-\alpha\, p_0 + 2\, \alpha^2\, p_0^2\right), 
& \hspace*{1cm} & p_{\nu}^2 = p_{\nu}^2 \left(1-2\, \alpha p_0 + 10\, \alpha^2\, p_0^2\right),
\eea
where $p_{0}$ is momentum at low energy. The parameter $\alpha=\alpha_0/(c\, M_{pl}) =\alpha_0 l_{pl}/\hbar$ \cite{afa1,afa2}, where $c$, $\alpha_0$, $M_{pl}\, (l_{pl})$ are speed of light as introduced by Lorentz and implemented in special relatively, dimensionless parameter of order one, and Planck mass (length), respectively. Then in comoving frame, the dispersion relation is given as
\bea
E_{\nu}^2 = p_{\nu}^2\, c^2 \left(1-2\,\alpha\,p_0\right) + M_{\nu}^2\, c^4.
\eea
When taking into consideration a linear dependence of $p$ on $\alpha$ and ignoring the higher orders of $\alpha$, then the Hamiltonian is
\bea \label{eq:Hh1}
\textbf{H} &=& \left(p_{\nu}^2\, c^2 - 2\, \alpha\, p_{\nu}^3 \, c^2 + M_{\nu}^2\, c^4 \right)^{1/2}.
\eea
The derivative of Eq. (\ref{eq:Hh1}) with respect to the momentum results in a comoving time-dependent velocity, i.e. a Hamilton equation, 
\bea
v(t) 
 &=& \frac{c}{a(t)} \left(1-2\alpha p_0 -\frac{M_{\nu}^2 c^2}{2 p_{\nu}^2} + 
\alpha p_0 \left[\frac{ M_{\nu}^2 c^2}{p_{\nu}^2} -
 \frac{ M_{\nu}^2 c^4}{p_{\nu}^2 c^2 + M_{\nu}^2 c^4} +
 \frac{ M_{\nu}^2 c^4}{p_{\nu}^2 c^2 + M_{\nu}^2 c^4} \frac{M_{\nu}^2 c^2}{2 p_{\nu}^2}\right]
\right). \label{eq:vt}
\eea 
The comoving momentum is related to the physical one through \hbox{$p_{\nu}=p_{\nu_{0}}(t_0)/a(t)$}, where $a$ is the scale factor, which in turn can be related to the redshift $z$
\bea \label{eq:at1}
a(z) &=& \frac{1}{1+z}.
\eea  
In the relativistic limit, $p\gg M$, the fourth and fifth terms in Eq. (\ref{eq:vt}) simply cancel each other. Then 
\bea
v(z) &=& c\,(1+z)\left[1-2\, \alpha\, (1+z)\, p_{\nu_0}  - \frac{M_{\nu}^2\, c^2}{2 (1+z)^2 p_{\nu_0}^2} +
\alpha\, \frac{ M_{\nu}^4 c^4}{2\, (1+z)^3\, p_{\nu_0}^3} 
\right]. \label{eq:vz}
\eea
In getting this expression, $p_0$ is treated as a comoving momentum. Then, it become straightforward to deduce the relative change in the relative velocity 
\bea \label{eq:dvz1}
\frac{\Delta v(z)}{c} &=& \alpha\,\left(- 2\, (1+z)^2\, p_{\nu_0} + \frac{ M_{\nu}^4 c^4}{2\, (1+z)^2\, p_{\nu_0}^3} \right) - \frac{M_{\nu}^2\, c^2}{2 (1+z) p_{\nu_0}^2}. \label{eq:vz2}
\eea
Despite the entire results will be given in section \ref{sec:mesure}, few remarks can be outlined here. 
The curves in left panel of Fig. \ref{fig:deltat1} represent the results of our approach. For a massless muon neutrino, the sign of  $\Delta v(z)/c$ remains negative with increasing $z$. When the muon neutrino mass is taken into account, the sign turns to positive. In this case, its value nearly vanishes at large $z$.  Accordingly, the resulting sign of the summation of first two terms of Eq. (\ref{eq:dvz1}) is determined by the second term, i.e. positive, at small $z$.  Then, the sign is flipped to negative at $z\sim 0.2$, i. e., the first term becomes dominant. 

The comoving redshift-dependent distance travelled by the particle of interest is defined as
\bea \label{eq:rz}
r(z) &=& \int_0^z \frac{v(z)}{(1+z)\, H(z)} dz,
\eea 
where $H(z)$ is the Hubble parameter depending on $z$. From Eqs. (\ref{eq:vz}) and (\ref{eq:rz}), the time of flight reads
\bea \label{eq:tnu}
t_{\nu} &=& \int_0^z \left[1-2\,\alpha\, (1+z)\, p_{\nu_0}  - \frac{M_{\nu}^2\, c^2}{2 (1+z)^2 p_{\nu_0}^2} +
\alpha\, \frac{ M_{\nu}^4 c^4}{2\, (1+z)^3\, p_{\nu_0}^3} \right] \frac{d z}{H(z)},
\eea
which counts for the well-known time of flight of a prompt low-energetic photon (first term). In other words, the time of flight is invariant in Lorentz symmetry. Furthermore, it is apparent that Eq. (\ref{eq:tnu}) contains a time of flight delay given as
\bea \label{eq:deltat1}
\Delta t_{\nu} &=&  \int_0^z \left[2\alpha \left( (1+z)\, p_{\nu_0} - \frac{ M_{\nu}^4 c^4}{4\, (1+z)^3\, p_{\nu_0}^3} \right) + \frac{M_{\nu}^2\, c^2}{2\, (1+z)^2 p_{\nu_0}^2}  \right] \frac{d z}{H(z)}.
\eea
It is clear that the first and second terms are due to LIV effects stemming from GUP. Both have $\alpha$ parameter. The third term reflects the effects of the particle mass on the time of flight delay. Furthermore, the second term alone seems to contain a mixed effects from LIV (GUP) and rest mass.

In order to determine $\Delta t_{\nu}$, Eq. (\ref{eq:deltat1}), it is essential to find out observational results and/or reliable theoretical model for the redshift-dependence of the Hubble parameter $H$. What we have is that $H$ depends on a time-dependent redshift, $d z/d t$,
\bea
H(z) &=& \frac{1}{a(z)}\, \left(\frac{d a(z)}{d z} \; \frac{d z}{d t}\right) = - \frac{1}{1+z}\, \frac{d z}{d t}.
\eea
It is obvious that this expression can be deduced from Eq. (\ref{eq:at1}). In general, the expansion rate of the Universe varies with the cosmological time \cite{Tawfik:2009mk,Tawfik:2009nh,Tawfik:2010bm,Tawfik:2010ht,Tawfik:2010pm,Tawfik:2011mw,Tawfik:2011sh,Tawfik:2011gh}. It depends on the background matter/radiation and its dynamics \cite{Tawfik:2011sh}. The cosmological constant reflecting among others the dark matter content seems to affect the temporal evolution of $H$ \cite{Tawfik:2011mw}.  Fortunately, the redshift $z$ itself can be measured with a high accuracy through measuring the spectroscopic redshifts of galaxies having certain uncertainties ($\sigma_z\leq 0.001$). Based on this, a differential measurement of time, $d t$, at a given redshift interval automatically provides a direct and clean measurement of $H(z)$ \cite{hz1,hz2,hz3}. These measurements can be used to derive constraints on essential cosmological parameters \cite{const}. In present work, we  implement the measurements of the expansion rate and their constrains in evaluating the integrals given in Eq. (\ref{eq:deltat1}).

\section{Confronting with Observations and Measurements}
\label{sec:mesure}

First we compare with recent observations of the early-type galaxies, which apparently provide a direct probe for the dependence of Hubble parameter $H$ and $z$. Making use of LIV contributions to $\Delta v/c$ and $\Delta t$, we study the dependence of each of these quantities on $z$ and compare the meanwhile-wrong-declared results with OPERA in section \ref{operra}. {\color{red} The lesson we gain from such a comparison is the ability of GUP even in judgement about edge-cutting observations.}
That distant neutrinos feel $z$-shift is discussed in section \ref{suboperra}. Then, the ultra high energy cosmic rays (UHECR) are utilized as a laboratory to study the consequences of LIV. Following the  $\gamma$-ray observations from Mrk $501$, constrains on Lorentz invariance breaking parameter based on potential departure from exact Lorentz invariance introduced in a  perturbative framework are motivating the comparison with UHECR.

\subsection{Early-type galaxies}
\label{earlyGal}

\begin{figure}[htb]
\includegraphics[angle=-90,width=8.cm]{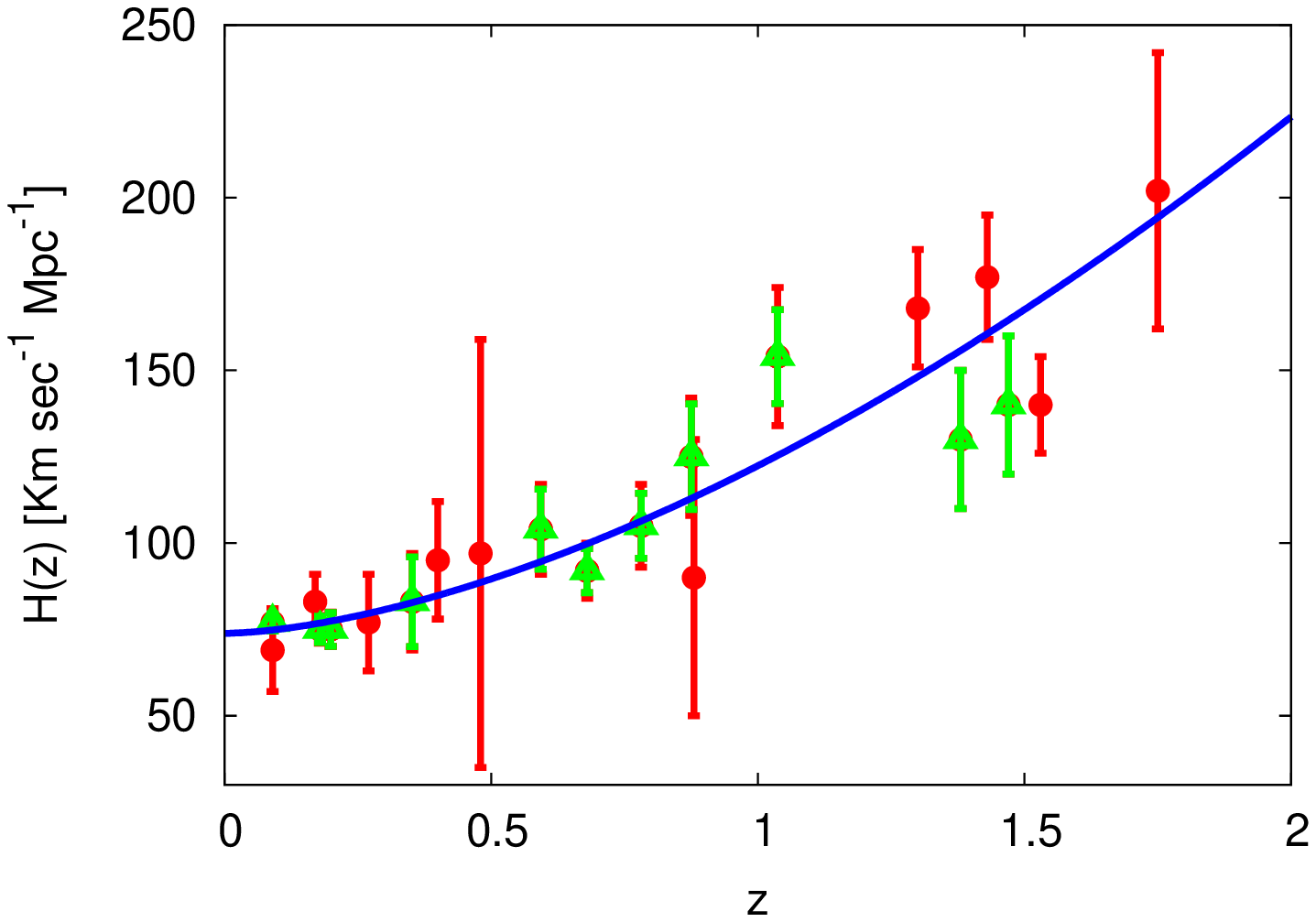}
\includegraphics[angle=-90,width=8.cm]{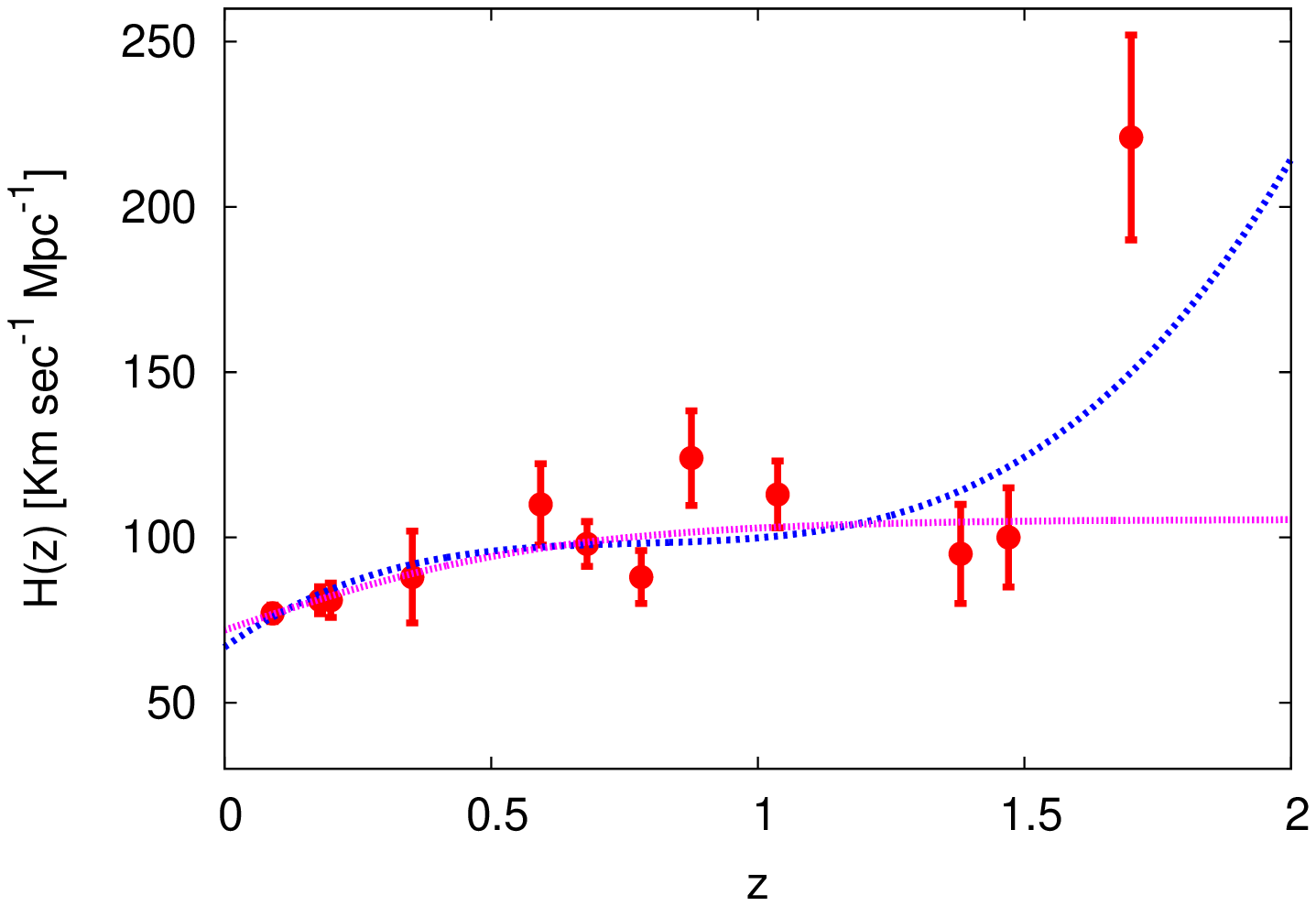}
\caption{Left-hand panel: Hubble parameter $H$ calculated from BC03 model (open triangle) and in combination with CMB data constraining possible deviations from standard (minimal) flat $\Lambda$CDM model (solid circles) is given in dependence on redshift $z$. The results from MS model are drawn in right-hand panel. The curves represent the fitting parameters (see text for details).
\label{fig:Hofz1} }
\end{figure}

Out of a large sample of early-type galaxies (about $11000$) extracted from several spectroscopic surveys spanning over $\sim 8\times 10^9$ years of cosmic look-back time, i.e. $0.15<z<1.42$ \cite{hz2}, most massive, red elliptical galaxies, passively evolving and without signature of ongoing star formation are picked up and used as standard cosmic chronometers \cite{const}.  The differential age evolution turns to be accessible, which gives an estimation for the cosmic time and can directly probe the dependence of Hubble parameter $H$ and $z$. A list of  new measurements of $H(z)$ with $5-12\%$ uncertainty is introduced in Ref. \cite{hz2}. The uncertainty in these observational data seems to be comparable with our calculation for $H(z\sim0.2)$. Figure \ref{fig:Hofz1} illustrates these observations as estimated in the BC03 \cite{bc03} model. They are combined with the cosmic microwave background (CMB) data and can be used to set constrains on possible deviations from the standard (minimal) flat $\Lambda$CDM model \cite{hz3}. The right panel shows a data set taken from MS model \cite{mastro}. It is obvious that the results are model-dependent.

The observational measurements can be fitted as follows. For the results obtained from BC03  model \cite{bc03} and using a combination with CMB data and setting constrains on possible deviations from the standard (minimal) flat $\Lambda$CDM model \cite{hz3}, the expression
\bea \label{eq:measr_hz1}
H(z)=\beta_1 + \gamma_1\, z + \delta_1\, z^2,
\eea
where $\beta_1=72.68\pm3.03$, $\gamma_1=19.14\pm 5.4$ and $\delta_1=29.71\pm 6.44$, fits well with the observations. The solid curve in left-hand panel of Fig. \ref{fig:Hofz1} represent the results from this expression. For the MS model \cite{mastro} measurements, we suggest two expressions
\bea 
H(z)&=&\beta_2 + \gamma_2\, z + \delta_2\, z^2 + \epsilon_2\, z^3, \label{eq:measr_hz2}\\
H(z)&=& \beta_3+\gamma_3\, \text{tanh}(\delta_3\, z), \label{eq:measr_hz3}
\eea
where $\beta_2=66.78\pm8.19$, $\gamma_2=113.27\pm7.5$, $\delta_2=-140.72\pm 12.6$, $\epsilon_2=60.61\pm5.48$, $\beta_3=71.94\pm4.35$, $\gamma_3=33.51\pm 7.94$ and $\delta_3=1.6\pm0.1$. The results of Eq. (\ref{eq:measr_hz2}) are given by dashed curve in the right-hand panel of Fig. \ref{fig:Hofz1}. Equation (\ref{eq:measr_hz3}) is drawn by dotted curve, where the largest point is excluded while renaming points build up the ensemble used in the fitting.
It is obvious that the implementation of Eq. (\ref{eq:measr_hz2}), which is obviously a rational function, in Eq. (\ref{eq:deltat1}) results in a non-analytic integral. On the other hand, implementing Eq. (\ref{eq:measr_hz3}) in Eq. (\ref{eq:deltat1}) makes the second and third integrals non-solvable. The first term can be solved, Appendix \ref{app:a}, where the results are also illustrated, graphically. 

It is apparent that Eq. (\ref{eq:measr_hz1}) simplifies the integrals given in Eq. (\ref{eq:deltat1}). Accordingly, there are  two types of LIV contributions to the time of flight delay. The first type is originated in finite $\alpha$. Finite $\alpha$ appears in two terms as follows.
\bea
2\alpha\, p_{\nu_0}\, \int_0^z  (1+z)\, \frac{d z}{H(z)} &=& \frac{\alpha}{\gamma}\,p_{\nu_0}\, \left[\ln\left[\beta_1+z(\gamma_1 + \delta_1 z)\right]-\frac{2(\gamma_1-2 \delta_1)}{A}\text{atan}\left(\frac{\gamma_1+2 \delta_1 z}{A}\right)\right], \hspace*{10mm}\label{eq:part1} \\
- 2\alpha\, \frac{ M_{\nu}^4 c^4}{4\, p_{\nu_0}^3} \int_0^z  \frac{1}{(1+z)^3}\, \frac{d z}{H(z)} &=&  \frac{-\alpha}{(\beta_1-\gamma_1+\delta_1)^3}\, \frac{ M_{\nu}^4 c^4}{4\, p_{\nu_0}^3}\, \left[\frac{2(\gamma_1-2 \delta_1)(\beta_1-\gamma_1+\delta_1)}{1+z}  \right. \nonumber \\ 
&+& \left. \left(3 \gamma_1 \delta_1 - \gamma_1^2 + \delta(\beta_1-3 \delta_1)\right)\ln\left(\beta_1+z(\gamma_1+\delta_1 z)\right)  \right. \label{eq:part2} \\
&-& \left. \frac{(\beta_1-\gamma_1+\delta_1)^2}{(1+z)^2} +2\left(\gamma_1^2 - 3 \gamma_1 \delta_1+\delta_1(3 \delta_1-\beta_1)\right)\ln(1+z) \right. \nonumber \\
&-& \left. \frac{2(\gamma_1-2\delta_1)}{A}\left(\gamma_1^2-\gamma_1 \delta_1+\delta_1(\delta_1-3 \beta_1)\right)\text{atan}\left(\frac{\gamma_1+2 \delta_1 z}{A}\right) \right], \nonumber 
\eea
where $A=(4 \beta_1 \delta_1 - \gamma_1^2)^{1/2}$. 
Furthermore, Eq. (\ref{eq:measr_hz1}) gives an exclusive estimation for the mass contribution to the time of flight delay,  
\bea
\frac{M_{\nu}^2\, c^2}{2\, p_{\nu_0}^2} \int_0^z\, \frac{1}{(1+z)^2}\, \frac{d z}{H(z)} &=& \frac{1}{(\beta_1-\gamma_1+\delta_1)^2}\, \frac{M_{\nu}^2\, c^2}{2\, p_{\nu_0}^2} \left\{\frac{\gamma_1^2-2 \gamma_1 \delta_1 + \delta_1 (\delta_1-\beta_1)}{A}\text{atan}\left(\frac{\gamma_1+2 \delta_1 z}{A}\right) \right. \nonumber \\ 
&-& \left. \frac{\beta_1-\gamma_1+\delta_1}{1+z} - \frac{1}{2}(\gamma_1-2 \delta_1)\ln\left[\frac{(1+z)^2} {\beta_1+z(\gamma_1+\delta_1 z)}\right] \right\}. \label{eq:part3}
\eea
The results are discussed in section \ref{operra}. Although the meanwhile-wrong-declared OPERA measurement on faster-than-light muon neutrino anomaly, $\Delta t$, and the relative change in the speed of neutrino $\Delta v$ in dependence on the redshift $z$ turn to be wrong, we utilize its main features to estimate $\Delta v$ and $\Delta t$.

\subsection{Comparing $\Delta t$ and $\Delta v$ with controversial OPERA Measurements }
\label{operra}
 
\begin{figure}[htb]
\includegraphics[angle=-90,width=8.cm]{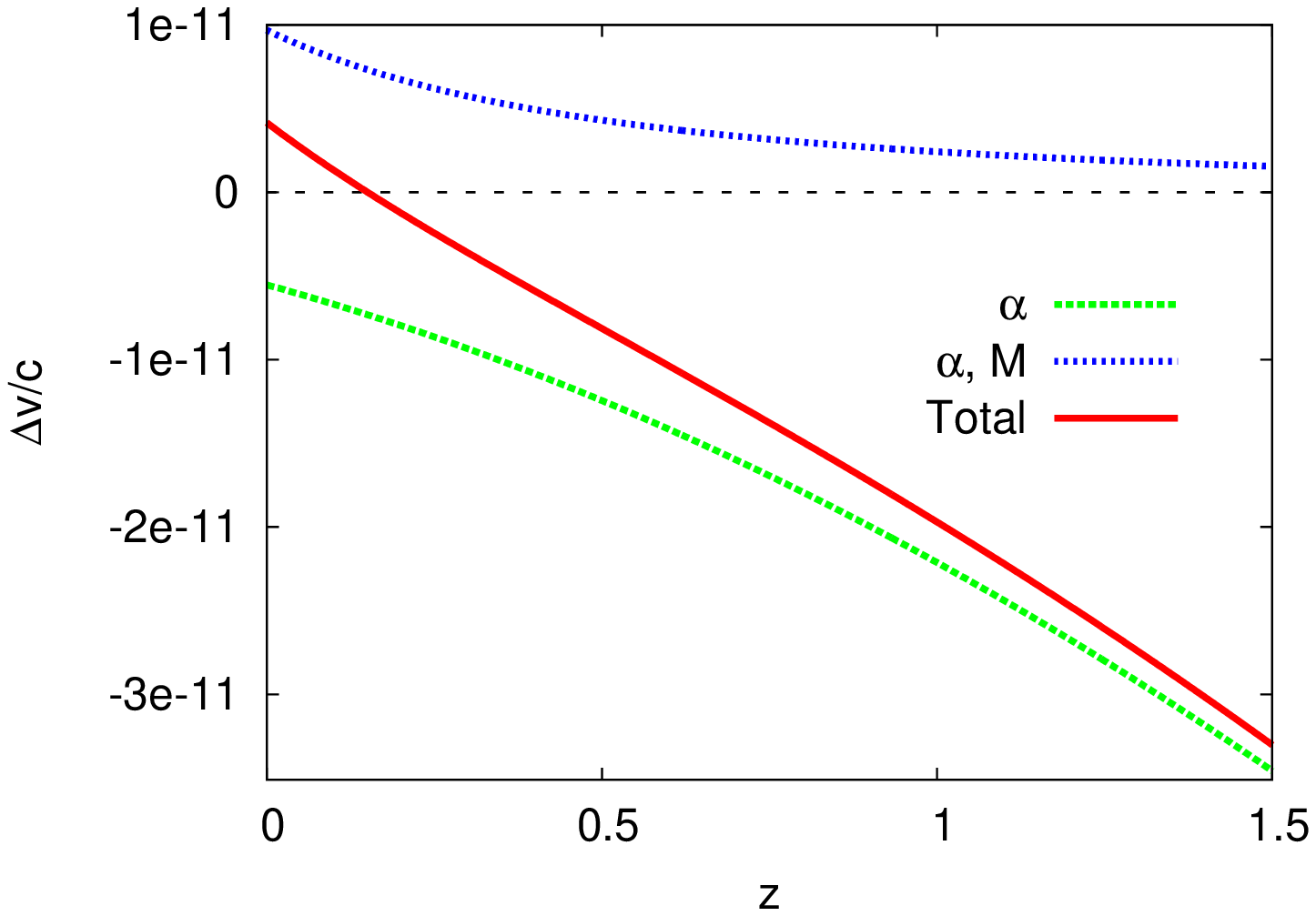}
\includegraphics[angle=-90,width=8.cm]{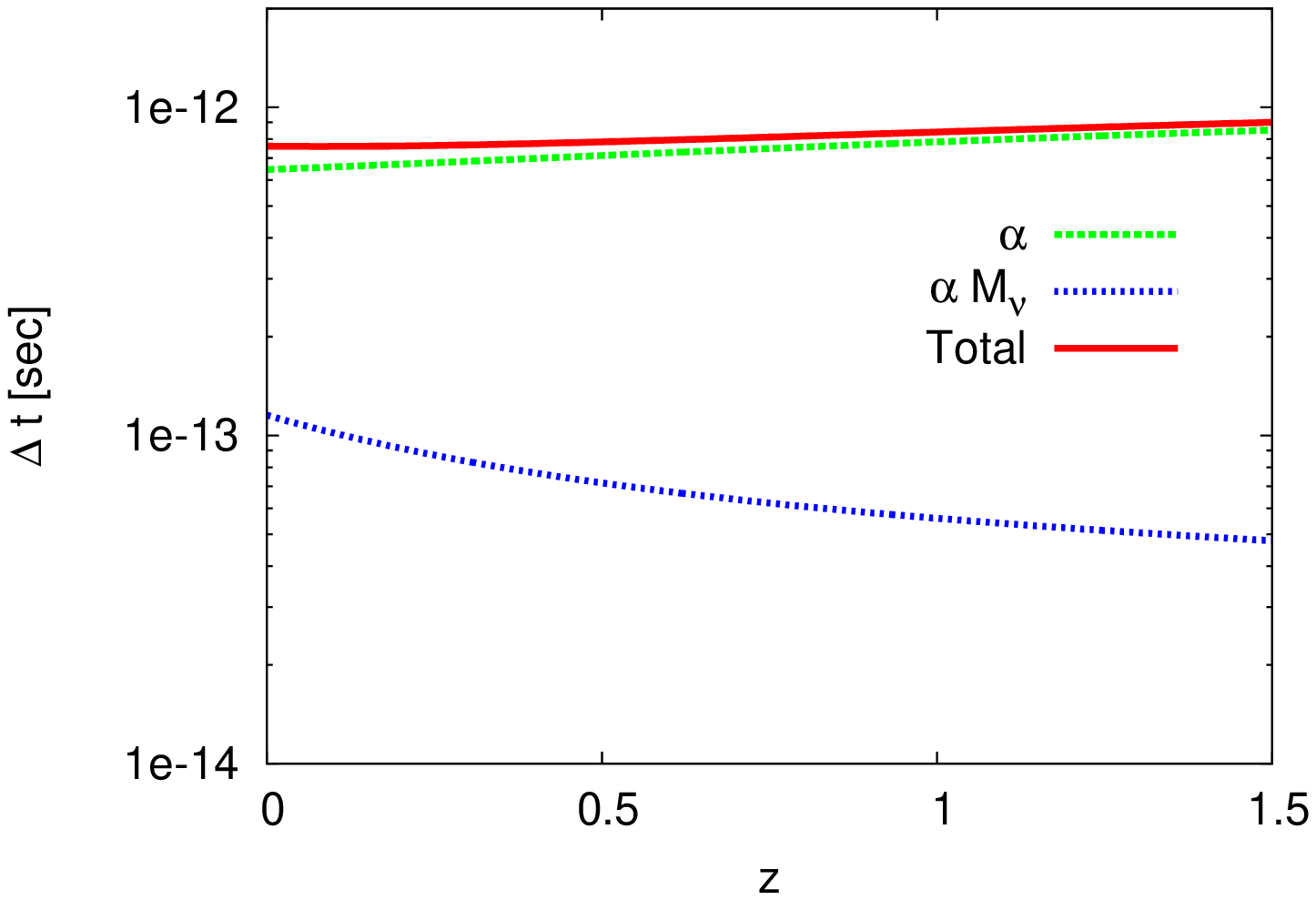}
\caption{Although the controversial OPERA measurement on faster-than-light muon neutrino anomaly, $\Delta t$, and the relative change in the speed of neutrino $\Delta v$ in dependence on the redshift $z$ turn to be wrong, using its main features, the relative change in the velocity of muon neutrino is given as a function of redshift $z$ in the left-hand panel. The right-hand panel shows the time of flight delay. The different curves represent different contributions to $\Delta v/c$ and $\Delta t$ (see text).
}
\label{fig:deltat1} 
\end{figure}

For the neutrino beam covering the distance between the source at CERN and the OPERA detector at the underground Gran Sasso Laboratory (LNGS), $\simeq 730\,$km, a time of flight delay of $\simeq 1045.1\pm11.3\,$nano seconds is first believed to be registered \cite{opera1}. 

As discussed in previous section, LIV comes up with two sources of contributions to $\Delta v/c$ and $\Delta t$. The first source is stemming from finite $\alpha$ and vanishing mass, Eq. (\ref{eq:part1}). The second source requires finite $\alpha$ and mass, Eq. (\ref{eq:part2}). The dependence of each of these quantities on $z$ are presented in left-hand panel of Fig. \ref{fig:deltat1}. In performing these calculations, we use the same parameters of the controversial OPERA experiment in which a faster-than-light muon neutrino anomaly has been claimed \cite{opera1,opera2}. They are the muon neutrino mass $M_{\nu}=1\,$eV and beam energy $E_{\nu}=17\,$GeV. The comparison with our approach assumes that the muon neutrino beam has a distant origin and was witnessing a huge redshift $z$ while the Universe expanded. Then, the time of flight delay $\Delta t$ can be calculated in dependence on redshift $z$. The first two terms of Eq. (\ref{eq:deltat1}) are calculated and drawn in right panel of Fig. \ref{fig:deltat1}. They are labelled by $\alpha$ and $\alpha, M_{\nu}$, respectively. We find that the first term, $\alpha$, is one or two orders of magnitude higher than the second one, $\alpha, M_{\nu}$. It is apparent that the sum of these two terms combines all LIV-sources. Accordingly, the time of flight delay can be approximated as $\Delta t \sim 10^{-12}\,$sec. 

Disregarding its confident and statistical interpretation, the comparison with the LIV time of flight delay leads to the conclusion that OPERA measurement is too huge (about six orders of magnitude) to be understood as LIV. 

Furthermore, the wrong OPERA experiment suggested an increase in the speed of light by $\sim7.5\,$km/sec ($\sim25$ part per millions $c$) \cite{opera1}. In a recent measurement \cite{opera2}, it is found that the difference between speed of muon neutrino and speed of light ranges from $-1.8$ to $2.3$ part per millionth $c$. The left-hand panel of Fig \ref{fig:deltat1} presents the redshift evolution of the possible change in the velocity of muon neutrino according to LIV. It is assumed that the mass of muon neutrino $M=1\,$eV and its energy $17\,$GeV. The first two terms of Eq. (\ref{eq:dvz1}) are compared with each other. 
The results are illustrated in left-hand panel of Fig \ref{fig:deltat1}. 

We notice that the first term (massless muon neutrino) results in a negative speed difference. Its absolute value increases almost linearly with increasing $z$. The second term remains in the positive site of the ordinate. The resulting speed difference is positive. While abscissa raises, the value of second term decreases, exponentially. The upper and lower values of speed difference range between $\sim 10^{-11}$ and $0$. With increasing $z$, the sum of these two terms changes the sign of $\Delta v(z)/c$. At small $z$, the second term seems to be dominant. At $z\sim 0.2$, the positive sign is switched into negative. At larger $z$-values, the first term becomes dominant. The average speed difference can be approximated as $\Delta v\sim -2\times 10^{-11}\, c$. Comparing to the value measured in OPERA \cite{opera1,opera2}, the LIV-value is about six orders of magnitude smaller. 

That the sign of $\Delta v(z)\equiv c_{\nu}-c$ is flipped meaning that
\bea \label{eq:cases}
\Delta v(z) &=& \left\{\begin{array}{l l} 
{\cal O}(+\text{ve}), & \text{then}\hspace*{5mm} c_{\nu}=c+ {\cal O}\\
 & \\
{\cal O}(-\text{ve}), & \text{then}\hspace*{5mm} c_{\nu}=c- {\cal O}\\
\end{array} \right.
\eea 
where $c_{\nu}$ is the velocity of muon neutrino. In Eq. (\ref{eq:cases}), the second case apparently follows the Lorentz invariance symmetry. The first case suggests that the speed of light would not be constant in all inertia frames. Furthermore, it would not be the maximum of travelling matter and information in the universe. The value of the additional quantity ${\cal O}$ is about $\sim 10^{-11}\, c$, i.e. ${\cal O}\sim 3\,$mm/sec, which indicates a superluminal propagation of high-energy muon neutrino at $z\lesssim 0.2$.

It was believed that OPERA gave results comparable to MINOS \cite{minos2}, where the value of the relative speed change was found as ${\cal O}\sim 10^{-5}$. On the other hand, these measurements are not compatible with the observations of $\sim 10\,$MeV-neutrino from supernova SN1987a \cite{supernova35}. In these observations, the value of ${\cal O}$ is estimated as $\sim 10^{-9}$. Therefore, the faster-than-light anomaly is energy-dependent. It drops rapidly, when reducing energy from GeV- to MeV-scale \cite{stecker1}. Nevertheless, the velocity anomaly is conjectured to reflect the propagation of all decay channels of neutrino and new physics such as LIV. 

{\color{red}
Few remarks on comparison with OPERA are now in order.
\begin{itemize}
\item The energy and mass of muon neutrino do not matter, as the applicability of GUP is not doubtable.
\item The wrong OPERA measurements are neither approved nor disapproved.
\item Our GUP approaches are not biased. Therefore, we present the comparison even after withdrawing OPERA measurements.
\item It intends to illustrate trust-able judgement about an even edge-cutting conclusion.
\end{itemize}
}

\subsubsection{Can distant neutrinos feel $z$-shift?}
\label{suboperra}.

Before CMB, the extremely long interaction length of neutrinos while traversing the relic background leads to integrate over cosmic time, or redshift, in order to estimate their survival probability \cite{neut1,neut3}.  This would be considered as an indirect observation that CMB-neutrinos would feel the redshift. According to standard cosmology, neutrinos should be the most abundant particles in the Universe, especially after CMB photons. Even, the CMB temperature can be expressed in dependence on redshift $z$, $T_{CMB}(z)=2.7 (1+z)~$K. While traversing the expanding Universe we live in, the effective relic UHECR neutrino density per unit redshift reads $n_{\nu 0}(1+z)/H(z)d z$ \cite{neut1,neut3}. The indirect dependence on $H(z)$ means that the possibility that the observation of neutrino absorption could even reveal the thermal history of the Universe, becomes high \cite{neut3}. Furthermore, the GRB neutrino flux in dependence on redshift can be estimated \cite{neut2}. 
Last but not least, we refer to the $\gamma$-ray observations from Mrk $501$ which assumes constrains on the Lorentz invariance breaking parameter based on potential departure from exact Lorentz invariance introduced in a  perturbative framework \cite{stecker1,cg1}. Accordingly, we could assume that the sensitivity of neutrinos to redshift might not be negligible. Amelino-Camelia {\it et al.} \cite{nature1} proposed to use astrophysical objects to look  for energy dependent time of arrival delays. As will be discussed in the section that follows, fixing the sensitivity factor and its energy dependence are essential inputs for this purpose. The sensitivity factor is related to the special time of flight delay and the time structure of the signal.  Furthermore, a weak indication for redshift dependence of time delays suggestive of LIV has been observed by Ellis {\it et al.} \cite{jellis1}. They  investigated the energy dependent time offsets in different energy bands on a sample of gamma ray bursts and, assuming standard cosmological model.

%
\subsection{Ultra High Energy Cosmic Rays}
\label{sec:uhecr}

Following from Eqs. (\ref{eq:tnu}) and (\ref{eq:deltat1}), the time of flight is conjectured to possess a delay of factor $\Delta t$. The generic ultra high energy cosmic rays (UHECR) can be utilized as a laboratory to study the consequences of LIV.  The pair production is kinematically allowed, when energy available to $\gamma$-rays \cite{stecker1} 
\bea
E_{\gamma}\geq m_e \left(\frac{2}{|{\cal D}|}\right)^{1/2}, 
\eea
where the subscript $e$ stands for electron or positron and ${\cal D}=(v_e-c)/c$. Depending on the values of $v_e$ and $c$, ${\cal D}$ can positive and negative. 

Stecker and Glashow used $\gamma$-ray observations from Mrk $501$ constraining the Lorentz invariance breaking parameter \cite{stecker1} based on potential departure from exact Lorentz invariance introduced in a  perturbative framework \cite{cg1}. According to Eq. (\ref{eq:cases}), we  can for simplicity assume that the electron has the same energy and mass as that of the muon neutrino in the OPERA experiment. 
The observations of UHECR refer to the existence of electrons with energies $E_{e}\simeq 1\times 10^{12}\,$eV \cite{nebula2} and $\gamma$-rays with energies $E_{\gamma}\geq 50\times 10^{12}\,$eV \cite{nebula} are observed. These observation would set upper limits to ${\cal D}_e \simeq 1.3\times 10^{-13}$ and ${\cal D}_{\gamma} \lesssim 2\times 10^{-16}$, respectively. It is apparent that all these values are smaller than the values that was estimated using the GUP-approach, $\Delta v\simeq 10^{-11}\, c$, section \ref{operra}. Such a discrepancy would be interpreted as follows. In our calculations, the GUP-approach assumes a {\it linear} momentum modification \cite{afa1,afa2}. As discussed above, this approach combine string theory, loop quantum gravity, black hole physics and doubly spacial relativity. 

Recent theoretical work on quantum gravity, especially within string theory, shows that the sensitivity factor of gamma ray bursts (GRB)  $\eta$ can be related to $\Delta t^*$, the special time of flight delay and $\delta t$, the time structure of the signal $\eta\equiv |\Delta t^*|/\delta t$ \cite{nature1}. The special time of flight delay is characterized by $E_{qg}\,$ ($E_{pl}$) effective quantum gravity energy scale (Planck energy scale). The condition that $E_{qg}\approx E_{pl}$, is that quantum gravity energy reaches the Planck energy. At this scale, $\eta$ is determined by $ |\Delta t^*|/\delta t$. In present work, $\Delta t^*$ is taken equivalent to $\Delta t$. Depending on distant origin, GRB emission can reach the Earth with different time structures $\delta t$. Therefore, the time structure might be sophisticated.  

On the other hand, the conventional gravitational lensing is achromatic. Therefore, the energy-dependent time delay would not be dependent on the actual emission mechanism GRB. Couple decades ago, lensed GRB was observed \cite{grblense}. It can be used to estimate the sensitivity factor as $\eta\approx 10^{-6}$. This value reveals that $\delta t \approx 10^{-7}\,$sec.  It is found that $\eta\approx 10^{-10}$ and therefore $\delta t \approx 10^{-3}\,$sec, when pulsars, supernovae and other astrophysical phenomena, but not GRB, are taken into consideration \cite{ref17}. A third estimation was done using neutrinos stemming from type-II supernovae, like SN1987a. In this case, $\eta\approx 10^{-4}$ and the time structure can be estimated as $\delta t\approx 10^{-9}\,$sec. 
In principle, the upper bound on $\alpha$ parameter which characterizes the GUP-approach can be found by comparing the calculations with the experimental observations \cite{afa2}. 

Confronting our calculations to UHECR requires fixing the sensitivity factor and its energy dependence.  The sensitivity factor is related to the special time of flight delay and the time structure of the signal. {\color{red} To judge about the applicability of GUP on UHECR, we recall the two main scenarios of their origin. Bottom-top scenario assumes that the cosmic rays are generated at low energies. Over their path to the Earth they gained energy through various mechanisms \cite{Tawfik:2010qf}. The top-bottom scenario proposes that the cosmic rays are produced at much higher energies (Planck scale). Over their path to the Earth they lost energies through various mechanisms \cite{Tawfik:2010qf}. Thus, the applicability is guaranteed.
}


\section{Conclusions and Outlook}
\label{sec:conl}

In this paper we introduced calculations for the time of flight delays and the relative change in the velocity of muon neutrino with mass $1\,$eV and energy $17\,$GeV. In doing this, we utilized the GUP-approach, which is based on a momentum-dependent modification in the standard dispersion relation. For a particle having a distant origin and energy comparable with the Planck energy scale, the co-moving momentum is given as a series of {\it linear} modifications on momentum. Varying redshift, we have calculated the relative change in the speed of massive muon neutrino and its time of flight delays. The redshift depends on the temporal evolution of the Hubble parameter, which can be estimated from a large sample of early-type galaxies extracted from several spectroscopic surveys spanning over $\sim 8\times 10^9$ years of cosmic lookback time, most massive, red elliptical galaxies, passively evolving and without signature of ongoing star formation are picked up and used as standard cosmic chronometers giving a cosmic time directly probe for $H(z)$. The measurements according to BC03 model and in combination with CMB data constraining the possible deviations from the standard (minimal) flat $\Lambda$CDM model are used to estimate the $z$-dependence of the Hubble parameter. The measurements based on MS model are used to show that the results are model-dependent.

We compared the results with the OPERA experiment. We conclude that the OPERA measurements for $\Delta t$ and $\Delta v$ are too large to be interpreted as LIV. Depending on the rest masses, the propagation of high-energy muon neutrino can be superluminal. The other possibility is not excluded. The comparison with UHECR reveals the potential discrepancy between an approach combining string theory, loop quantum gravity, black hole physics and doubly spacial relativity and a perturbative departure from exact Lorentz invariance.  For reliable confronting of our calculations to UHECR, we need to fix the sensitivity factor and its energy dependence. The sensitivity factor is related to the special time of flight delay and the time structure of the signal.

In light of this study, we believe that GRB would be able to set an upper bound to the GUP-charactering parameter $\alpha$. Furthermore, the velocity anomaly is conjectured to reflect the propagation of all decay channels of neutrino and new physics such as LIV.

\section*{Acknowledgements}

This work of AT and HM is partly supported by the German-Egyptian Scientific Projects (GESP ID: 1378).

\appendix
\section{Time of Flight Delay According to MS Model}
\label{app:a}

\begin{figure}[htb]
\includegraphics[angle=0,width=8.cm]{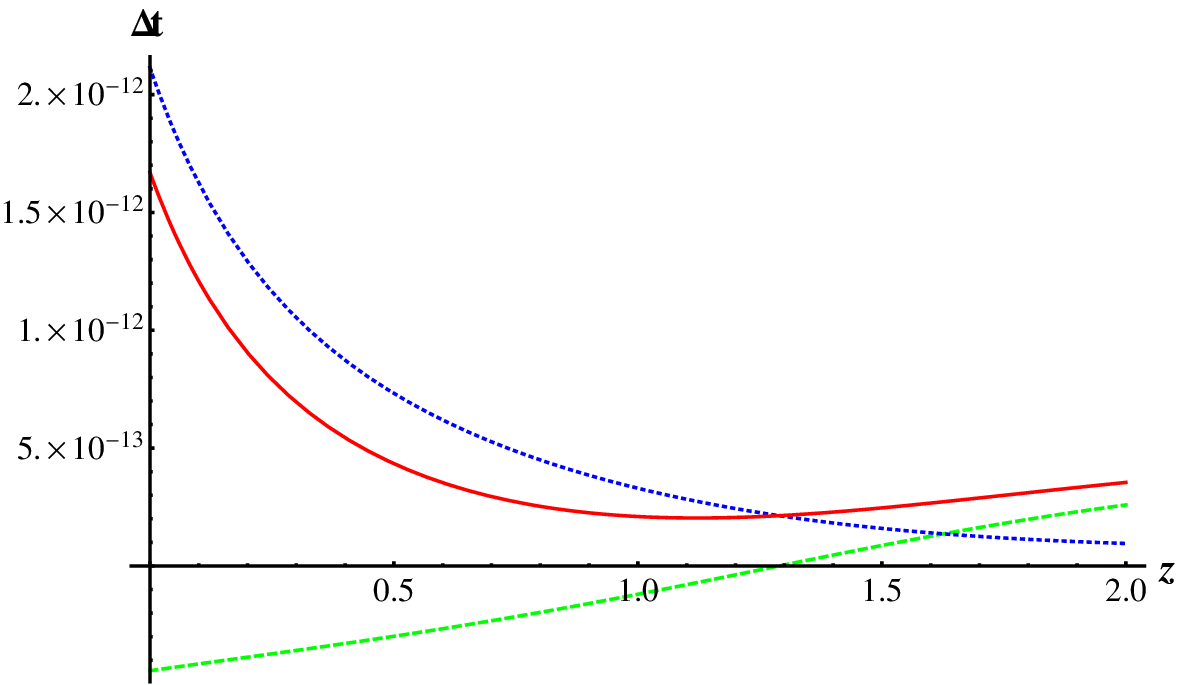}
\includegraphics[angle=0,width=8.cm]{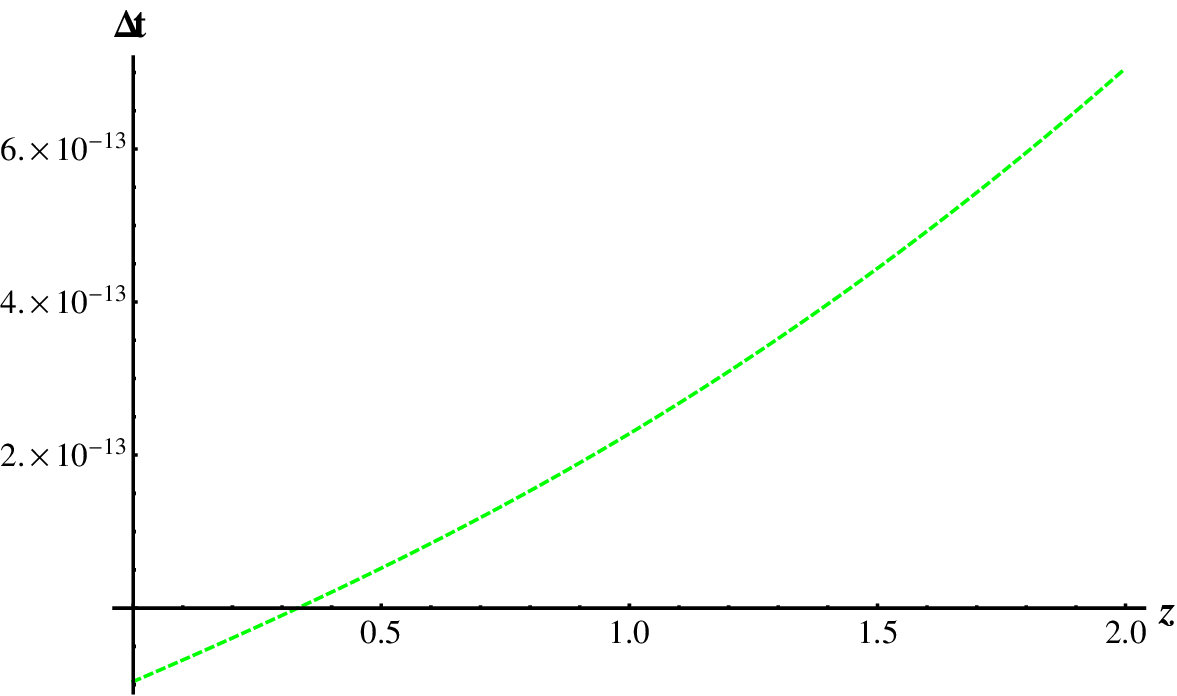}
\caption{Left panel: the time of flight delay of muon neutrino (mass $1\,$eV and energy $17\,$GeV) in dependence on $z$, using (\ref{eq:measr_hz2}) in Eq. (\ref{eq:deltat1}). The first term is given by dashed curve, while dotted curve represents the second term. The right panel draws Eq. (\ref{eq:a1}). Both dashed curves seem to represent comparable results. 
}
\label{fig:deltat4} 
\end{figure}

It is apparent that integrating the rational expression (\ref{eq:measr_hz2}) into Eq. (\ref{eq:deltat1}) gives a numerical solution. In left panel of Fig. \ref{fig:deltat4}, the first (dashed curve) and second (dotted curve) terms of Eq. (\ref{eq:deltat1}), where $H(z)$ is taken from (\ref{eq:measr_hz2}), are given in dependence on $z$. Their summation is given by the solid curve. The time of flight delay, $\Delta t$ can be averaged as $\sim 10^{-13}\,$sec. This value is much smaller than the one measured in OPERA experiment, so that the latter would not interpreted by LIV.

When implementing Eq. (\ref{eq:measr_hz3}) into Eq. (\ref{eq:deltat1}), the integrals in the second and third terms can not be solved, analytically. The first term can be solved as follows.
\bea 
\Delta t(z) &=& 2\alpha\, p_{\nu_0}\, \int_0^z  (1+z)\, \frac{d z}{H(z)} \nonumber \\
&=&   \frac{\alpha\, p_{\nu_0}\; B\, \text{sech}\left(\delta_3\, z \right) }{e^{C}\,\beta_3 \left(\beta_3^2-\gamma_3^2\right) \delta_3^2 \left[\beta_3+\gamma_3 \text{tanh}\left(\delta _3\, z \right)\right]}
 \left\{ 
\beta_3^2\, \delta_3^2\, e^{C} z (2+z)  +  
\left[ \left(
\sqrt{1-\frac{\beta_3^2}{\gamma_3^2}} - e^{C} \right) \gamma_3^2\, \delta_3^2\, z^2   \right. \right. \label{eq:a1} \\
&+& \left. \left. \beta _3 \gamma_3 e^{C} \left(
i\, \pi  \ln\left[1+e^{2 \delta _3 z}\right] - 2\, C \ln\left(1-e^{\delta_3\, z - 2 C}\right) 
- i\, \pi  \ln\left[\text{cosh}\left(\delta_3\, z \right)\right]+ 
 2\, C \ln\left[i\, \text{sinh}\left(C + \delta_3\, z \right)\right] \right. \right. \right. \nonumber \\
&-& \left.  \left.  \left. 2\, \delta_3 \left\{ \frac{\pi}{2}  z + z\, C + z\, \ln\left(1-e^{-2 \left[C + \delta _3\, z \right]}\right) +
  \ln B
\right\}
\right)
\right] 
\right\}, \nonumber 
\eea
where $B=\beta _3 \text{cosh}\left(\delta_3\, z \right) + \gamma _3 \text{sinh}\left(\delta _3\, z \right)$ and $C=\text{atanh}\left(\beta_3/\gamma_3\right)$. The results are drawn in the right panel of Fig. \ref{fig:deltat4}. In these calculations, only the real component of the second line of Eq. (\ref{eq:a1}) is taken into consideration. The values of $\Delta t$ can be approximated to $10^{-13}\,$sec, which about seven orders of magnitude smaller than the time of flight delay measured in OPERA experiment. 
With the dashed curve (first term) in left panel, this term gives comparable results, qualitatively and almost quantitatively.

{\color{red}
\section{Bounds on GUP parameter}

The GUP parameter is given as $\alpha=\alpha _0/(M_{p} c)=\alpha _0 \ell _p/\hbar$, where $c,\; \hbar$ and $M_p$ are  speed of light and Planck constant and mass, respectively. The Planck length $\ell _p\, \approx\, 10^{-35}~$m and the Planck energy $M_p c^2 \,\approx \, 10^{19}~$ GeV.  $\alpha _0$, the proportionality constant, is conjectured to be dimensionless \cite{afa1}. In natural units $c=\hbar=1$, $\alpha$ will be in GeV$^{-1}$, while in the physical units, $\alpha$ should be in GeV$^{-1}$ times $c$. The bounds on $\alpha_0$, which was summarized in Ref. \cite{afa2,Chemissany:2011nq,dvcjp}, should be a subject of precise astronomical observations, for instance gamma ray bursts. 
\begin{itemize}
\item Other alternatives were provided by the tunnelling current in scanning tunnelling microscope and the potential barrier problem \cite{Ali:2010yn}, where the energy of the electron beam is close to the Fermi level. We found that the varying tunnelling current relative to its initial value is shifted due to the GUP effect \cite{Chemissany:2011nq,Ali:2010yn}, $\delta I/I_0\approx 2.7 \times 10^{-35}$ times $\alpha _{0} ^{2}\,$. 
In case of electric current density $J$ relative to the wave function $\Psi$, the current accuracy of precision measurements reaches the level of $10^{-5}$. Thus, the upper bound $\alpha_0<10^{17}$. Apparently, $\alpha$ tends to order $10^{-2}~$GeV$^{-1}$ in natural units or $10^{-2}~$GeV$^{-1}$ times $c$ in physical units. This quantum-mechanically-derived bound is consistent with the one at the electroweak scale \cite{Chemissany:2011nq,Ali:2010yn,dvcjp}. Therefore, this could signal an intermediate length scale between the electroweak and the Planck scales  \cite{Chemissany:2011nq,Ali:2010yn,dvcjp}.
\item On the other hand, for a particle with mass $m$ mass, electric charge $e$ affected by a constant magnetic field ${\vec B}=B_{\hat z}\approx10~$Tesla, vector potential ${\vec A}= B\,x \,{\hat y}$ and cyclotron frequency $\omega_c = eB/m$, the Landau energy is shifted due to the GUP effect \cite{Chemissany:2011nq,Ali:2010yn} by
\bea
\frac{\Delta E_{n(GUP)}}{E_n} &=& -\sqrt{8\, m}\; \alpha\;  (\hbar\, \omega _c )^{\frac{1}{2}} \,
 \left(n +\frac{1}{2}\right)^{\frac{1}{2}}  \approx  - 10^{-27}\; \alpha_0.
\eea
Thus, we conclude that if $\alpha_0\sim 1$, then $\Delta E_{n(GUP)}/E_n$ is too tiny to be measured. But with the current measurement accuracy of $1$ in $10^3$, the upper bound on $\alpha_0<10^{24}$ leads to $\alpha=10^{-5}$ in natural units or $\alpha=10^{-5}$ times $c$ in the physical units.

\item Similarly, for the Hydrogen atom with Hamiltonian $H=H_0+H_1$, where standard Hamiltonian $H_0=p_0^2/(2m) - k/r$ and the first perturbation Hamiltonian $H_1 = -\alpha\, p_0^3/m$, it can be shown that the GUP effect on the Lamb Shift \cite{Chemissany:2011nq,Ali:2010yn} reads
\bea
\frac{\Delta E_{n(GUP)}}{\Delta E_n} &\approx & 10^{-24}~\alpha_0.
\eea
Again, if $\alpha_0 \sim 1$,  then $\Delta E_{n(GUP)}/E_n$ is too small to be measured, while the current measurement accuracy gives $10^{12}$. Thus, we assume that  $\alpha_0>10^{-10}$.
\end{itemize}

In light of this discussion, should we assume that the dimensionless $\alpha_0$ has the order of unity in natural units, then $\alpha$ equals to the Planck length $\approx\, 10^{-35}~$m. The current experiments seem not be able to register discreteness smaller than about $10^{-3}$-th fm, $\approx\, 10^{-18}~$m \cite{Chemissany:2011nq,Ali:2010yn}. We conclude that the assumption that $\alpha_0\sim 1$ seems to contradict various observations and  experiments \cite{Chemissany:2011nq,Ali:2010yn}. Therefore, such an assumption should be relaxed to meet the accuracy of the given experiments. Accordingly, the lower bounds on $\alpha$ ranges from $10^{-10}$ to $10^{-2}~$GeV$^{-1}$. This means that $\alpha_0$ ranges between $10^9\, c$ to $10^{17}\, c$. 

}



\end{document}